\documentstyle[aps,epsfig]{revtex}

\begin{document}
\draft

\title{Charge stripes seen with x-rays in $\bf
La_{1.45}Nd_{0.4}Sr_{0.15}CuO_4$}

\author{T. Niem\"oller, H. H\"unnefeld, J. R. Schneider}
\address{Hamburger Synchrotronstrahlungslabor HASYLAB at Deutsches
Elektronen-Synchrotron DESY\\
Notkestr. 85, D-22603 Hamburg, Germany}

\author{N. Ichikawa, S. Uchida}
\address{Department of Superconductivity, The University of Tokyo, Bunkyo-ku,\\
         Tokyo 113-8656, Japan}

\author{T. Frello, N.H.~Andersen}
\address{Condensed Matter Physics and Chemistry Department,
Ris\o\ National Laboratory,
DK-4000 Roskilde, Denmark}
\author{J. M. Tranquada}
\address{Brookhaven National Laboratory, Upton, NY 11973, USA}

%\date{\today}
\maketitle
%\widetext

\begin{abstract}
Superstructure reflections due to the ordering of holes into stripes
in $\rm La_{1.45}Nd_{0.4}Sr_{0.15}CuO_4$ have been studied with high energy
x-ray diffraction. These reflections have been observed clearly for the
first time
in a sample which is superconducting at low temperatures ($T_c \sim 10$~K).
The stripe peaks vanish above 62(5)~K whereas the magnetic signal of the
stripe ordering which has been seen with
neutrons before is already
suppressed at $\sim45$~K. Our results confirm that
the ordering of spins and holes is driven by the charges as it is found in
the case of La$_{1.6-x}$Nd$_{0.4}$Sr$_{x}$CuO$_4$ at the doping level
of $x=0.12$.
\end{abstract}

\newpage

\section{Introduction}

The ordering of holes and spins into stripes in the CuO$_2$-planes of
Nd-doped La$_{2-x}$Sr$_{x}$CuO$_4$ (LSCO) has attracted much attention, because
charge separation and antiferromagnetic spin fluctuations are believed to
be important
for understanding the mechanism of high temperature superconductivity.
The idea of pinned stripes in the low temperature tetragonal (LTT) phase
of Nd-doped LSCO or La$_{2-x}$Ba$_{x}$CuO$_4$ \cite{john_nature,axe89} provides
an explanation for the anomaly observed at $x \sim \frac{1}{8}$, where
superconductivity is destroyed, or at least  strongly suppressed
\cite{mood88,craw91,buch94,mood97}. The tilting of the oxygen octahedra
along the $[1\,0\,0]$ and $[0\,1\,0]$ directions (i.e., parallel to Cu--O
bonds) in
the LTT phase introduces a pinning potential for horizontal and vertical
stripes,
whereas in the low temperature orthorhombic (LTO) phase it is absent due to the
rotation of the tilting axis into the $[1\,1\,0]$ direction.

In a single crystal of La$_{1.48}$Nd$_{0.4}$Sr$_{0.12}$CuO$_4$, neutron
diffraction
allowed the observation of both magnetic and charge-ordering superstructure
reflections \cite{john_nature}. Consistent with the idea of pinning by the LTT
lattice modulation, static stripe ordering within the CuO$_2$ planes
appears at the
transition temperature from the LTO to the LTT phase, which is $\rm \sim 68 K$.
From the positions of the superlattice peaks and the nominal hole
concentration it
follows that stripes of holes are approximately half-filled, and act as
antiphase domain walls with respect to the antiferromagnetically ordered Cu
spins.  Thus, a Sr doping level of $\sim\frac{1}{8}$ yields a spacing between
stripes of $4a$, where $a$ is the lattice constant.
High energy x-ray studies have been succesful in confirming the results for the
charge stripe ordering in an $x=0.12$ sample \cite{martin98}.

It is still an open question as to whether charge
stripes are limited to hole concentrations near $\frac{1}{8}$ in Nd-doped
LSCO or
whether they influence the physics of cuprate superconductivity in general,
but there is accumulating evidence for the latter.
Inelastic neutron scattering experiments indicate the possibility of
moving, fluctuating stripes in LSCO \cite{cheo91,yamada98} and even in YBCO
\cite{dai98,mook98}.   Local charge ordering in
La$_{2-x}$Sr$_{x}$CuO$_4$ with $x\leq\frac18$ is suggested by a recent
nuclear quadrupole resonance (NQR) study \cite{hunt99}.
It is interesting to imagine that dynamic stripe correlations might be
necessary for
superconductivity \cite{emer97}, whereas pinning the stripes leads to a
strongly
reduced critical temperature $\rm T_c$, but the experimental evidence for
such a
scenario is still incomplete.
So far the Nd-doped LSCO system offers a unique
opportunity to study pinned stripe patterns in diffraction experiments,
giving information about the nature of charge and spin ordering.

The superstructure reflections due to the ordering of holes are shifted by
$2\epsilon$ in $h$- or $k$-direction relative to fundamental
reflections, whereas the magnetic peaks are located around the
antiferromagnetic peak position $(\frac12,\frac12,0)$, shifted by
$\epsilon$ also in $h$- or $k$-direction.
In neutron scattering experiments, the splitting $\epsilon$ has been
observed to
increase slightly as the Sr content increases from 0.12 to 0.15 and 0.20,
implying
a decrease in the average stripe spacing; however, it has only been
practical to
study the superstructure reflections due to the antiferromagnetic ordering
\cite{john97}.  It is now imperative to directly characterize the charge
order at
Sr concentrations away from $\frac18$.

In this work we present x-ray studies with 120~keV photons on
$\rm La_{1.45}Nd_{0.4}Sr_{0.15}CuO_4$. Our results are in complete
agreement with the stripe model and complementary to the
experimental evidence of
antiferromagnetic stripe ordering in this sample \cite{john97}.
As observed in the $x=0.12$ case \cite{john_nature}, the superstructure
reflections due to the ordering of the holes sets in at a higher temperature
compared to the magnetic signal, indicating that the transition into
the stripe phase is driven by the charge separation \cite{zach98}.

\section{Experiment}

The experiments have been performed on the triple-axis diffractometer
designed for the use of $\sim 100$~keV photons
at the high-field wiggler beamline
BW5 at HASYLAB, Hamburg \cite{bouchard98}.
X-ray diffraction in this energy range has proven to be very
successful in studying charge ordering in cuprates, nickelates and manganates
\cite{martin98,titi97,thomas99}.
As in neutron scattering experiments, the large penetration depth
($\sim 1$ mm) allows one to probe the bulk of the sample, enabling a direct
comparison of x-ray and neutron diffraction data.
In contrast to former experiments on $\rm La_{1.48}Nd_{0.4}Sr_{0.12}CuO_4$
\cite{martin98}, the Si/TaSi$_2$ monochromator and analyzer crystals
have been replaced
by the new Si$_{1-x}$Ge$_x$ gradient crystal material. These crystals show
very high reflectivity values of 96\%\ (not corrected for absorption) and
variable widths of the rocking curves depending on the Ge content
\cite{steffen98,steffen99}.
By use of this new material,
the scattered intensity at the stripe peak positions
in Nd doped LSCO is about four times higher
compared to the results obtained previously with the
utilization of $\rm Si/TaSi_2$
crystals as monochromator and analyzer \cite{steffen98user}.
Figure~1 shows a comparison of a superstructure reflection measured with the
$\rm Si/TaSi_2$ and Si$_{1-x}$Ge$_x$ gradient crystals. The count rate
collected with the gradient crystals is nearly two times higher, and a
slightly better signal to background ratio could be reached by reducing the
beam spot on the sample, i.e. by probing only the center of the crystal.
The lower curve has
been measured with the $\rm Si/TaSi_2$ crystals, illuminating a two times
larger sample volume.

In this experiment 120~keV photons have been employed, with a monochromatic
beam intensity of $1.2\times 10^{11}$ photons/$\rm mm^2$.
The resolution (FWHM) at a (2,0,0) reflection of
$\rm La_{1.45}Nd_{0.4}Sr_{0.15}CuO_4$ was 0.020~\AA$^{-1}$ in the
longitudinal and 0.0014~\AA$^{-1}$ in the transverse direction, the latter
being
limited by the sample mosaicity. At smaller diffraction angles the longitudinal
resolution improves because the diffraction geometry becomes less
dispersive, e.g. the FWHM of the (1,1,0) reflection is
0.011~\AA$^{-1}$.
A closed-cycle cryostat has been utilised and temperatures between 9~K and
300~K could be reached at the sample position. The studied crystal
($\sim2\times2\times4$ mm$^3$) is a piece of a cylindrical rod that was
grown by the travelling-solvent floating-zone method. 

\section{Results}

Previous hard x-ray diffraction experiments on a
$\rm La_{1.48}\-Nd_{0.4}Sr_{0.12}CuO_4$ crystal have
shown that the stripe peaks are displaced not only within the
$(h,k,0)$ plane, but also in the $\ell$ direction \cite{martin98}.
Due to the better resolution in reciprocal space compared to neutron
scattering experiments it has been possible to
find a modulated intensity of the stripe peaks
along the $\ell$-direction with maxima at
$\ell = \pm 0.5$, indicating a correlation of the charge stripes in
next-nearest-neighbor layers along the $c$-axis.
The highest scattering intensities for the stripe peaks
are expected at the positions $(2-2\epsilon,0,\pm0.5)$ and
$(2+2\epsilon,0,\pm0.5)$. Figure 2 clarifies the positions of the
stripe peaks in reciprocal space relative to the CuO$_2$ planes. We have used
the tetragonal unit cell with $a=b=3.775$~\AA, $c=13.10$~\AA.

Longitudinal scans along $(h,0,0.5)$ are shown in Fig.~3.  As anticipated,
small superstructure reflections are observed at $h=2\pm2\epsilon$ with
$2\epsilon = 0.256(1)$.  The peaks that are present at $T=9$~K have disappeared
after raising the temperature to 70~K.  The signal to
background ratio is $\sim 0.2$ for the $2-2\epsilon$ reflection and $\sim 0.1$
at the $2+2\epsilon$ position.
In both scans the background rises due to the vicinity of the (2,0,0)
reflection.  At this fundamental Bragg reflection
the count rate in the peak maximum is about $10^8$ times larger
than in the stripe peaks.

The curves through the data points in Fig.~3 are least-squares fits.  In
both (a)
and (b) the charge-order peaks are modelled with a Gaussian.  The
background in (a)
is approximated by the tail of a second Gaussian centered near $h=2$ plus a
linear
contribution, while only a linear background is used in (b).
Usually, the tail of a
fundamental reflection is Lorentzian shaped, but nevertheless it is
possible that locally other functions are a better approximation to a
non-linear background, since the shape of the
background is related to the sample
quality. The studied $\rm La_2CuO_4$ crystal with Sr and Nd as dopants
on the La-site certainly incorporates defects and strain which are
responsible for the rather high background, which is $\sim 150$ photons
per second at the superlattice positions.

In the inset of Fig.~3(a), two more low-temperature longitudinal scans,
slightly shifted in the $k$-direction, are displayed. The absence of any
peaks in
these scans indicates that the peak found at $(2-2\epsilon,0,0.5)$
is narrow in $k$.  This conclusion is confirmed by the 9~K transverse
scan shown in Fig.~4. The background in the transverse scans is linear, and a
fit of the stripe peak with a Lorentzian is slightly more successful
than utilizing a Gaussian. From the present data, it is difficult to
determine the true peak shape, given the small signal-to-background ratio.

Figure~5(a) shows the temperature variation of the amplitudes of the $h$ and
$k$ scans at the $(2-2\epsilon,0,0.5)$ position.
With rising temperature the amplitudes decrease rather linearly; the
peak vanishes at 62(5)~K, which is $\sim8$~K below the structural
transition from
the LTT into the LTO phase. The intensity of the (3,0,0) reflection,
also shown in Fig.~5(a), is a measure of this
transition because it only occurs in the LTT phase.
From the longitudinal and transverse scans a FWHM of 0.028(5)~r.l.u.
can be inferred, resulting in a correlation length of 43(8)~{\AA}.
The change of the peak widths with temperature can be seen
in Fig.~5(b). Up to $\sim55$~K it is rather constant, but above this
temperature the width of the peak in $h$- and $k$-directions increases
considerably. Note that the increasing width as well as the
vanishing peak intensity are not directly connected to the
structural transition from the LTT into the LTO phase, but occur at
a remarkably lower temperature.

\section{Discussion}

Our x-ray results provide direct evidence for charge-stripe order in
$\rm La_{1.45}Nd_{0.4}Sr_{0.15}CuO_4$.  Such charge order had previously been
inferred from the magnetic order
observed by neutron scattering \cite{john97}.  This is an important result,
because
high-field magnetization \cite{oste97} and muon-spin-rotation ($\mu$SR)
\cite{muon98} studies have indicated that bulk superconductivity also exists in
this sample.  Another conclusion of the $\mu$SR study is that static
magnetic order
is present throughout essentially the entire sample volume.  Thus, it
appears that
stripe order and superconductivity coexist intimately at $x=0.15$.

Compared to the $x=0.12$ composition, both the charge ordering temperature
and the
superstructure peak intensities are reduced.  The superlattice intensities
[normalized to the (200) peak] are roughly 30\%\ weaker in $x=0.15$ relative to
$x=0.12$.  A similar reduction in magnetic ordering temperature and
intensity have
been observed previously.  At both Sr concentrations, charge ordering
appears at a
higher temperature than magnetic order.

Within the error bars the correlation length of
the stripe order in the $x=0.15$ sample is the same as in the $x=0.12$ sample,
and in both samples the correlation length of the stripes is constant over
a large
temperature range. In contrast to this finding, it is important to note that
neutron scattering and muon-spin rotation experiments revealed that the
spins are
only quasi-static, with a decreasing correlation length, in the temperature
range
above 30~K in the $x=0.12$ sample \cite{muon98,john98}.
This continuous loss of correlation is only observed in
the magnetic signal, whereas
the correlation length for the charge order shows a decrease only very
close to the
disordering temperature. The temperature range in which a residual signal
with an
increased width is observed can be interpreted with fluctuating stripes,
since the detected signal in our x-ray diffraction experiment
is a sum of elastic and inelastic contributions.

One might conclude that fluctuations of stripes set in if the
pinning potential in the LTT phase
is destroyed at the structural transition
from LTT to LTO, as observed in the x~=~0.12 sample,
or if rising temperature competes with a pinning potential via
thermal activation.
In the sample with $x=0.15$, the second possibility holds,
since the suppression of the stripe ordering is not connected
to the structural transition. The pinning potential is smaller in the
$x=0.15$ sample than in the $x=0.12$ sample because
the tilting angle of the oxygen octahedra
is reduced with increasing Sr content in Nd doped LSCO \cite{buch94}.

On the other hand, the idea that the LTT structure is required for
charge-stripe
order is challenged by the interpretation of the Cu NQR results in LSCO by Hunt
{\it et al.} \cite{hunt99}.  That work suggests that static stripes can
occur even
within the LTO phase, and that the ordering temperature continues to
increase as
$x$ decreases below 0.12.  As an initial test of this picture, we intend to
study
charge order in a Nd-doped crystal with $x=0.10$ in the near future.

To summarize, we have been successful in validating the existence of stripe
order in
$\rm La_{1.45}Nd_{0.4}Sr_{0.15}CuO_4$, a sample which is superconducting
below $\sim 10$~K. Charge-order peaks due to stripe ordering have been observed
at $(2-2\epsilon,0,\pm0.5)$, $(2+2\epsilon,0,0.5)$, and at
$(0,2-2\epsilon,\pm0.5)$. The $2\epsilon$ value of
0.256(1) is in very good agreement with neutron scattering results for the
magnetic peaks. This is a further step in establishing the picture
of static stripes in Nd doped LSCO.
The intensity of the stripe signal decreases almost linearly with
rising temperature and vanishes at 62(5)~K, $\sim8$~K below the structural
transition from the LTT to the LTO phase.
The formation of the stripe pattern in
$\rm La_{1.45}Nd_{0.4}Sr_{0.15}CuO_4$ is less pronounced than in
$\rm La_{1.48}Nd_{0.4}Sr_{0.12}CuO_4$.
\\

Acknowledgements:\\
Work at Brookhaven is supported by
Contract No.\ DE-AC02-98CH10886, Division of Materials Sciences, U.S.
Department of Energy.

\begin{figure}
\caption{A comparison of charge peaks at the (2-2$\epsilon$,0,0.5) position
measured with $\rm Si/TaSi_2$ and $\rm Si_{1-x}Ge_x$ gradient crystals as
monochromator and anlyzer material. The new gradient materials give higher
primary
beam intensity. The signal to background ratio was improved by reducing the
beam
size and focusing on the center of the studied sample.}
\end{figure}

\begin{figure}
\caption{Based on the choice of lattice parameters shown in (a), the
positions of
the expected charge-order supperlattice peaks with the $(h,k,0.5)$ plane of
reciprocal space are indicated in (b). The unit cell at low temperatures is
of the
size $3.775\times 3.775\times 13.10$ \AA$^3$. The performed scans in reciprocal
space are indicated by short lines in (b).}
\end{figure}

\begin{figure}
\caption{Scans along $(h,0,0.5)$ through the superlattice peaks at (a)
$h=2-2\epsilon$ and (b) $h=2+2\epsilon$, at several temperatures.  The curves
through data points are least squares fits, described in the text. The
inset in (a)
shows similar scans performed about positions displaced slightly in $k$.}
\end{figure}

\begin{figure}
\caption{Transverse ($k$) scans through $(2-2\epsilon,0,0.5)$ at two different
temperatures.  The solid line is a Lorentzian on a linear
background.  The inset shows similar scans, but displaced in $h$, also
measured at
9~K.}
\end{figure}

\begin{figure}
\caption{(a) The amplitudes of the charge peaks in $h$ (longitudinal) and $k$
(transvers) scans at $(2-2\epsilon,0,0.5)$ decrease almost linearly with
temperature
and vanish at 62(5)~K. The intensity of the (3,0,0) reflection
(a measure for the LTT phase) is given by the filled circles. The LTT to LTO
transition temperature is 70(2)~K.  (b) The Gaussian width of the peak
measured along
both $h$ and $k$ as a function of temperature. A
broadening of the peak is observed above 55~K.}
\end{figure}

\clearpage

\setcounter{figure}{0}

\begin{figure}
\begin{center}
\epsfig{file=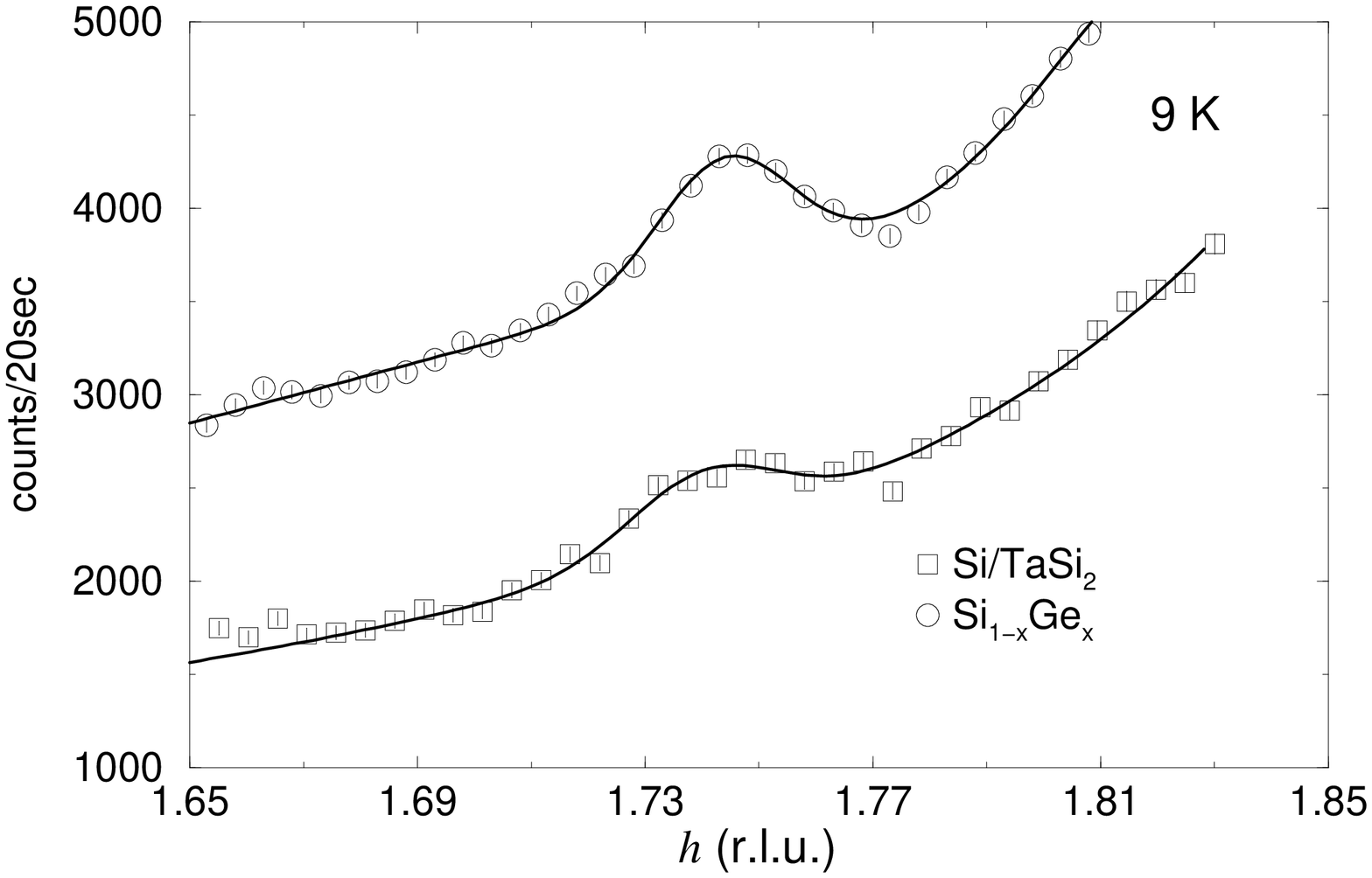,width=7cm,angle=-90}
\caption{}
\end{center}
\end{figure}

\begin{figure}
\begin{center}
\epsfig{file=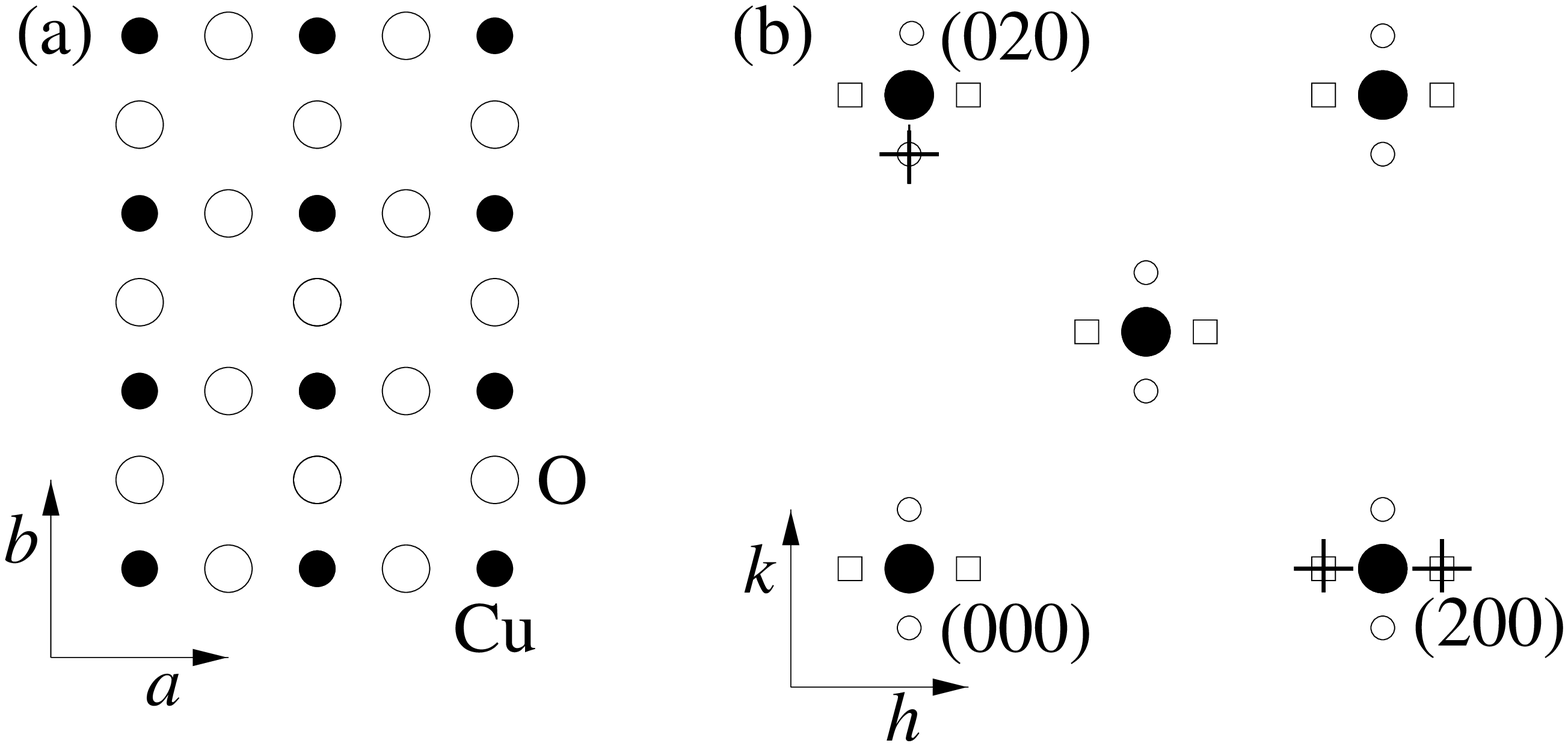,width=7cm,angle=0}
\caption{}
\end{center}
\end{figure}

\clearpage

\begin{figure}
\begin{center}
\epsfig{file=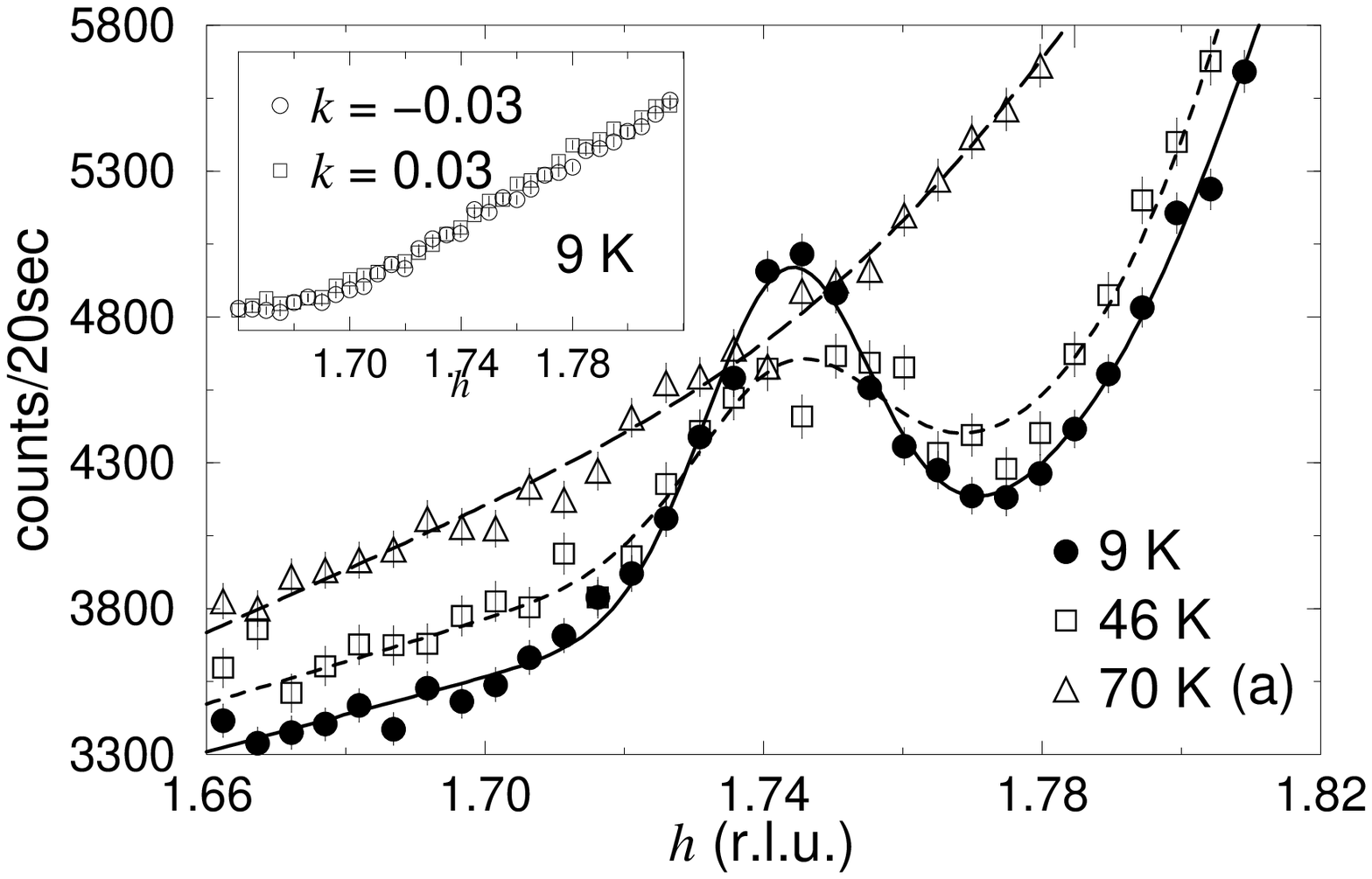,width=9cm,angle=0}
\caption{a}
\end{center}
\end{figure}

\setcounter{figure}{2}

\begin{figure}
\begin{center}
\epsfig{file=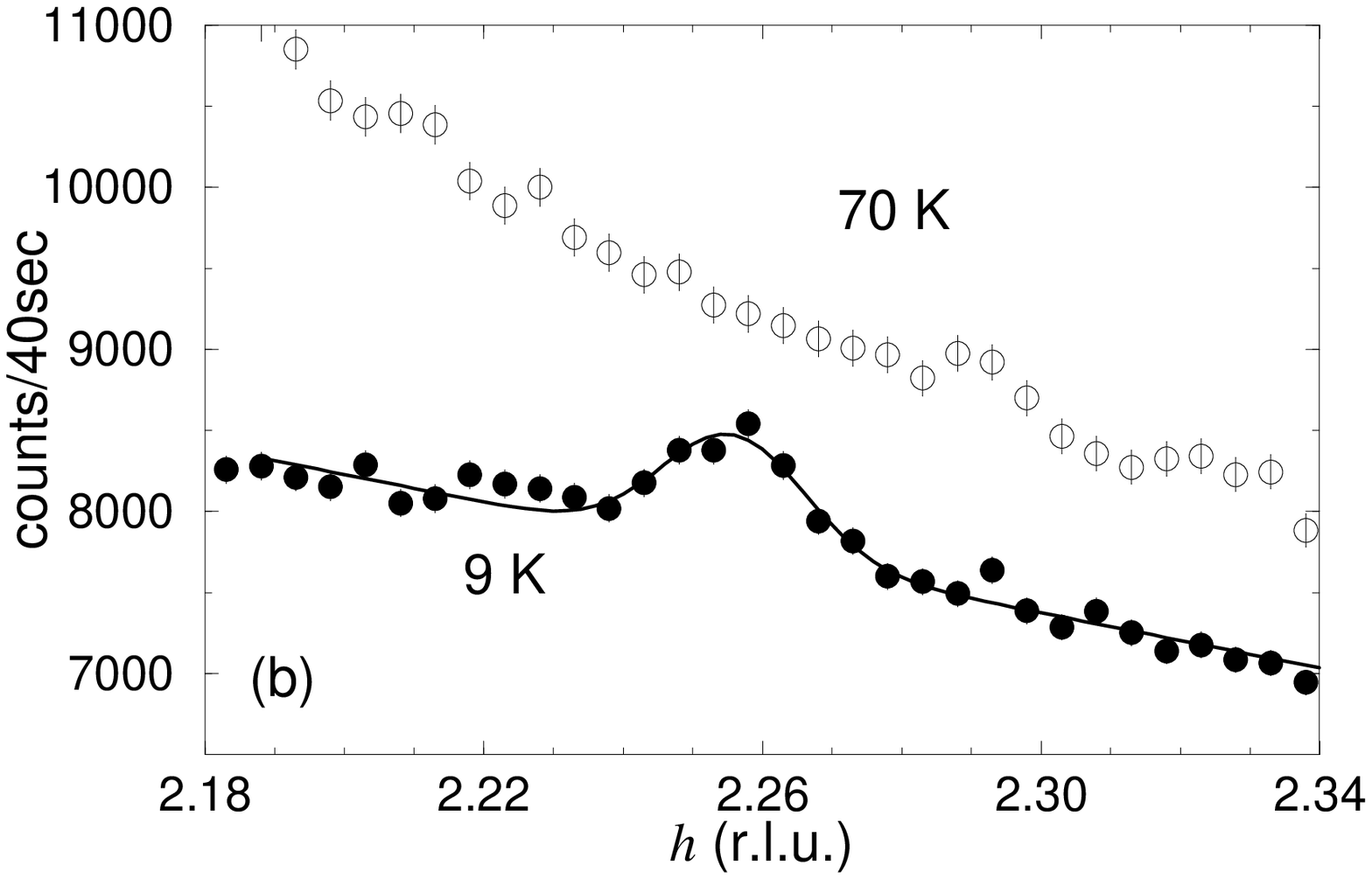,width=9cm,angle=0}
\caption{b}
\end{center}
\end{figure}

\clearpage

\begin{figure}
\begin{center}
\epsfig{file=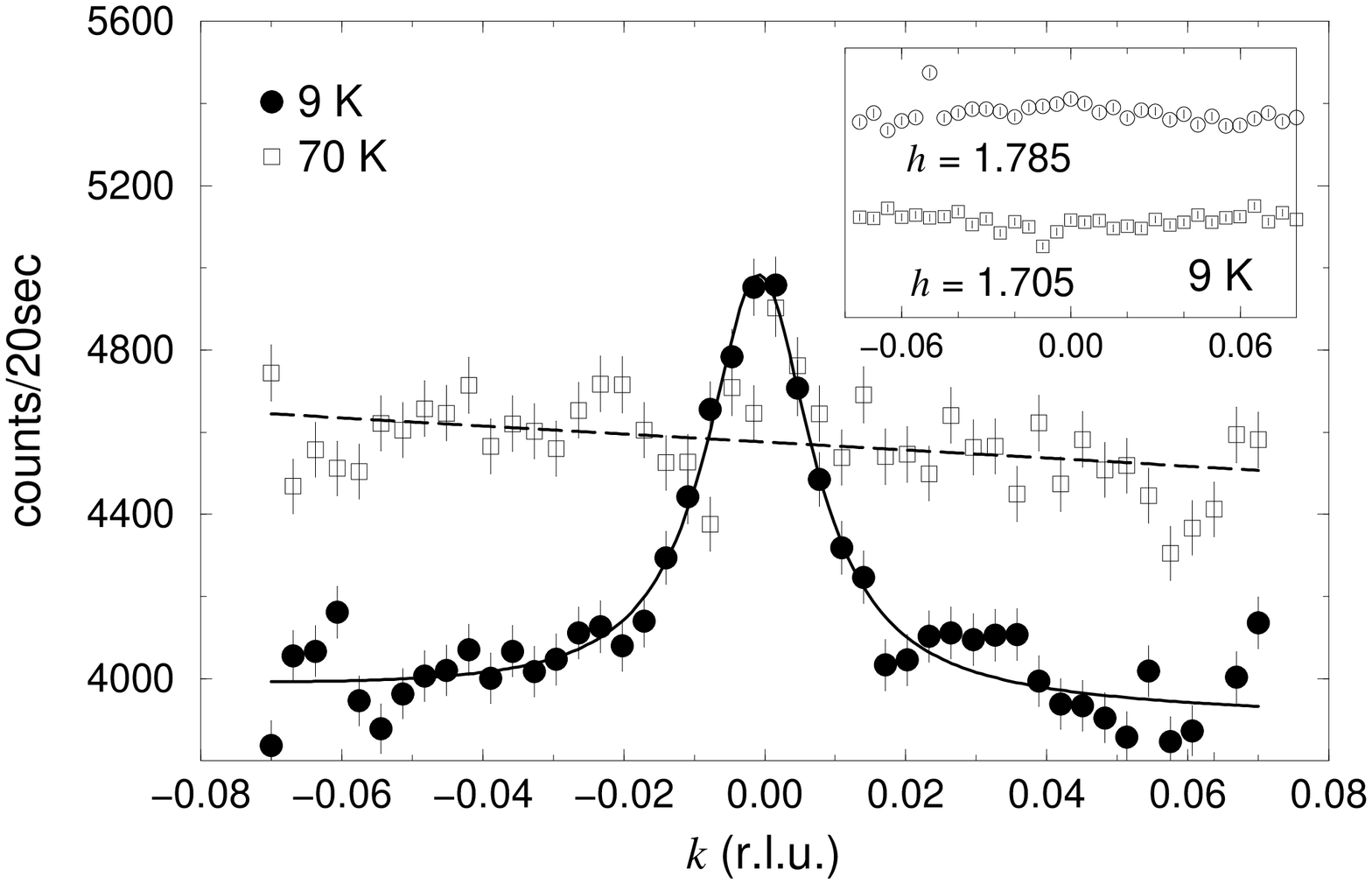,width=7cm,angle=-90}
\caption{}
\end{center}
\end{figure}

\begin{figure}
\begin{center}
\epsfig{file=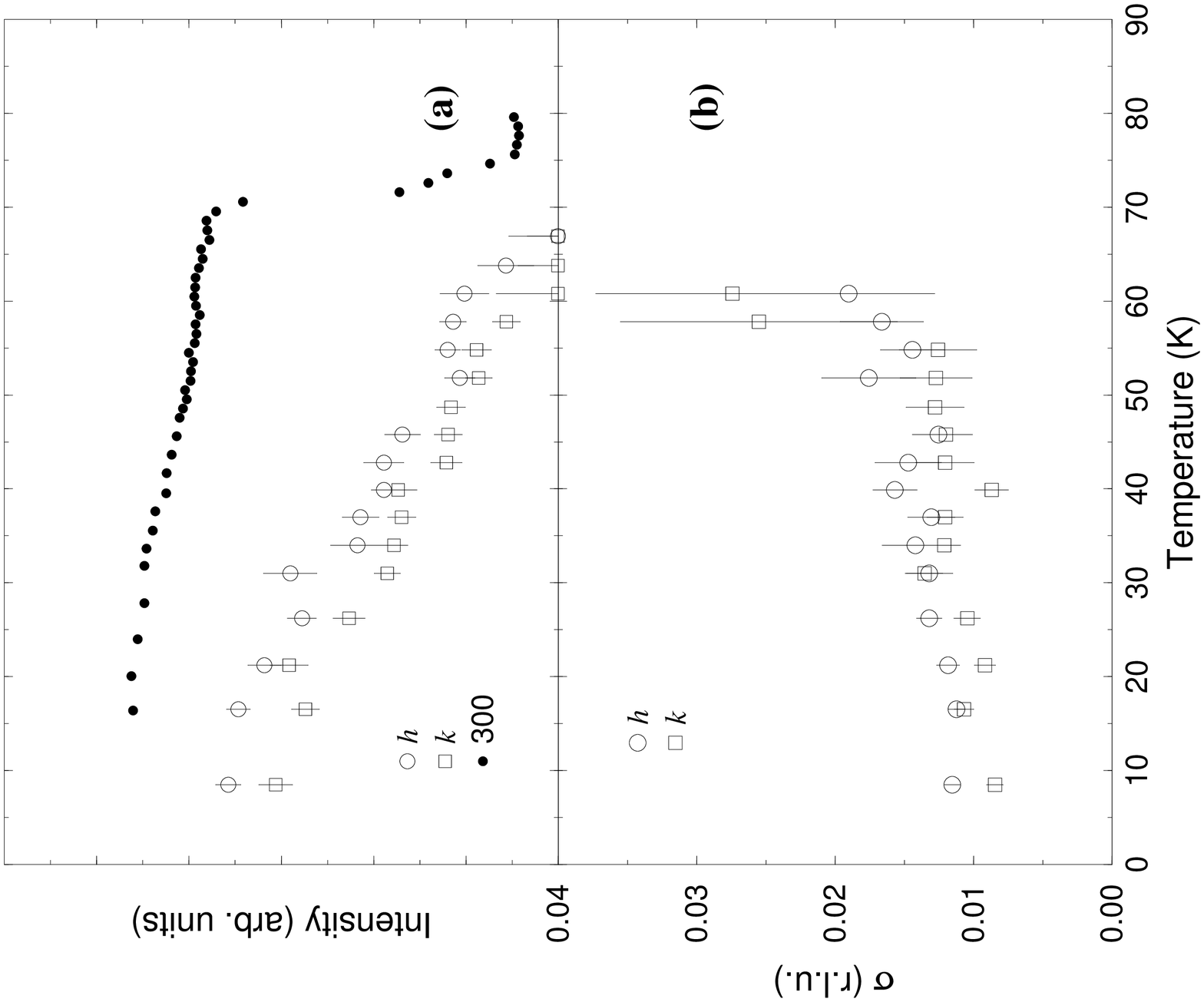,width=9cm,angle=-90}
\caption{}
\end{center}
\end{figure}

\end{document}